\documentclass[prd,aps,twocolumn,showpacs]{revtex4}

\usepackage{epsfig}
\usepackage{float}

\newcommand{\be}{\begin{equation}}
\newcommand{\ee}{\end{equation}}
\newcommand{\bea}{\begin{eqnarray}}
\newcommand{\beas}{\begin{eqnarray*}}
\newcommand{\eea}{\end{eqnarray}}
\newcommand{\eeas}{\end{eqnarray*}}
\newcommand{\ba}{\begin{array}}
\newcommand{\ea}{\end{array}}

\newcommand{\lra}{\leftrightarrow}
\def\ls{\mathrel{\lower4pt\vbox{\lineskip=0pt\baselineskip=0pt
           \hbox{$<$}\hbox{$\sim$}}}}
\def\gs{\mathrel{\lower4pt\vbox{\lineskip=0pt\baselineskip=0pt
           \hbox{$>$}\hbox{$\sim$}}}}

\begin{document}


\title{Left-right symmetric model with $\mu\leftrightarrow\tau$ symmetry.}
\author{Juan Carlos G\'omez-Izquierdo\footnote{jcarlos@fis.cinvestav.mx}
and Abdel
P\'erez-Lorenzana\footnote{aplorenz@fis.cinvestav.mx} }

\affiliation{Departamento de F\'{\i}sica, Centro
de Investigaci\'on y de Estudios Avanzados del I.P.N.\\
Apdo. Post. 14-740, 07000, M\'exico, D.F., M\'exico}

\date{May 2010}

\begin{abstract}
We analyze the leptonic sector in the left-right symmetric model dressed
with a $(Z_{2})^{3}$ discrete symmetry which realizes, after weak
spontaneous breaking, a small broken $\mu\lra\tau$ symmetry that
is suggested to explain observable neutrino oscillation data.
$\mu\lra\tau$ symmetry is broken at tree level in
the effective neutrino mass matrix due to the mass difference
$\widetilde{m}_{\tau}\neq \widetilde{m}_{\mu}$ in the diagonal Dirac mass terms, whereas all
lepton mixings arise from a Majorana mass matrix. In the limit
of a small breaking we
determined $\theta_{13}$, and the deviation from the maximal value of
$\theta_{23}$, in terms of the light neutrino hierarchy scale, $m_{3}$,
and a single free parameter $h_{s}$ of the model.
\end{abstract}

\pacs{14.60.Pq,12.60.-i,11.30.Fs}

\maketitle

\section{Introduction}
The standard model (SM) has been a successful theory that explains most particle
physics up to energies of about 100 $GeV$. However, as it is well known,
it provides no explanation for the phenomenon of neutrino
oscillations that gives clear indications for the existence of non zero
neutrino masses and mixings.
Tiny squared neutrino mass differences had been confirmed by a number of experiments,
including Kamiokande, Super-Kamiokande, KamLAND, SNO, K2K, and  Minos, among
others. All known data are well understood in the setup  where
there are three mass eigenstates $\nu_{1,2,3}$, and three weak
eigenstates $\nu_{e,\mu,\tau}$, which are related to each other through the
mixing matrix $U_{PMNS}$~\cite{pmns}. This matrix, in the standard
parametrization, has three mixing angles named $\theta_{23}$, $\theta_{12}$, and
$\theta_{13}$, one CP-violating Dirac phase, $\varphi$, and two Majorana
phases. The mixing angles $\theta_{23}$ and $\theta_{12}$ are measured in
atmospheric and solar neutrino experiments, respectively, with very good accuracy, whereas the third
angle, $\theta_{13}$, and the CP phases have not been measured yet,
although the first is known to be small.
Global data analysis on Ref.~\cite{valle1} provides
$\sin^2\theta_{12}=0.304^{+ 0.022}_{-0.016}$,
$\sin^2\theta_{23}=0.50^{+ 0.07}_{-0.06}$,
and
$\sin^2\theta_{13}=0.01^{+ 0.016}_{-0.011}$,
which is still consistent with zero ( see also Ref.~\cite{theta13} for an independent analysis regarding $\theta_{13}$ ).
A measurement of $\theta_{13}$ could be possible in future experiments~\cite{nextexp}, though.
Data also indicate that there are two scales for the squared mass differences,
$\Delta m_{\odot}^{2}= (7.65^{+0.23}_{-0.20})\times 10^{-5}~{\rm eV}^2$
and
$\Delta m_{ATM}^{2}= (2.4^{+0.12}_{-0.11})\times 10^{-3}~{\rm eV}^2$,
that define the oscillation lengths at any given energy.
All these prove that neutrinos are massive, and therefore that the SM needs to be extended to include their mass.

The simplest route to include neutrino masses and
mixings to the SM is to add the missing right-handed neutrino (RHN)
states  to the matter content, and then invoking the see-saw mechanism~\cite{see-saw}.
However, in this approach right the RHN mass scale is introduced by hand, with no relation whatsoever to the Higgs mechanism that gives mass to all other fields.
The  see-saw mechanism, on the other hand, comes in rather naturally in the context of
left-right symmetric extensions of the SM~\cite{leftright},
aside from other nice features, as for instance the recovery of parity symmetry, and the appearance of right-handed currents at high energy,
which makes such extensions also very appealing.
Nonetheless, in any of these simple matters or gauge extensions to the SM, a complete understanding of the peculiar mixing pattern of neutrinos is not possible without additional assumptions.
This is usually taken as a motivation to include  flavor symmetries to the models.
As a matter of fact, a theoretical understanding of such numbers has been the goal of many
theoretical works~\cite{mt} in the last several years among which the
bimaximal and tribimaximal mixing scenarios have attracted some
special attention in the literature.
Of particular interest in some of those models is the fact that mixing angles are consistent with $\theta_{23}\approx 45^{\circ}$,
$\theta_{13}\approx0^{\circ}$, which, in the fermion basis where charged lepton masses are diagonal, seems  to favor a $\mu\lra\tau$ symmetry in the neutrino mass matrix.
Indeed, a fundamental $\mu\lra\tau$ symmetry in the theory would exactly predict $\theta_{23}=45^{\circ}$, and  $\theta_{13}=0^{\circ}$.

Although $\mu\lra\tau$ symmetry is indeed a flavor symmetry in SM interactions, it is not an exact one for the whole particle physics due to the different masses of muon and tau leptons. Nevertheless,  having two of the predicted mixing angles so close to the observed values could be a good indication that $\mu\lra\tau$ symmetry does have something to do with fixing the observed lepton mixings at some level.
As it was shown in a recent analysis~\cite{juancarlos}, the effect of the breaking produced by the sole charged lepton sector on neutrino mixings is rather negligible. This is because in a SM with three right-handed neutrinos, dressed with an effective $\mu\lra\tau$ symmetry in all neutrino couplings, the $\mu$ and $\tau$ mass difference enters into the mixings only through one-loop quantum corrections mediated by the $W$ boson, and thus, it  comes out very suppressed. Besides, the implementation of the $\mu\lra\tau$ symmetry in the SM turns out to be quite subtle, since RHN's  do not  lie along  similar representations as other leptons. RHN's are rather singlets of the theory, which drives us into an ambiguous definition of the meaning of lepton flavor in this sector. As a consequence, there is no unique or natural way to  associate the properties of RHN under $\mu\lra\tau$ symmetry. Only two generic classes of models do actually exist, depending on whether or not RHN's carry lepton number with them, though.

The situation is expected to be different in the left-right symmetric model (LRSM),
based on $SU(2)_{L}\times SU(2)_{R}\times U(1)_{B-L}$ gauge
symmetry, with explicit parity symmetry.
LRSM has some advantages over the SM, in
particular, as we already commented, it includes automatically right-handed neutrino fields
in its matter content, providing a working frame where left-right handed fields are treated on the same footing. This is
particularly important to give meaning to the mass scale of the right-handed neutrinos, which here is now linked to the breaking of parity, as
well as to the scale of right-handed weak interactions. LRSM is also known to allow for Dirac and Majorana mass terms for neutrinos, and so, the see-saw mechanism naturally arises from it.
A common draw back of the model is that it introduces  many additional free parameters which have to be constrained. However currently this is hard to do due to the lack of experimental data.
All the phenomenology of LRSM is so far only constrained by SM data. An extended analysis of the LRSM parameters has been done in \cite{boris}.
Flavor symmetries in the context of LRSM had also been previously explored (see for instance~\cite{modmohapa}) and, as is usual on flavor extensions, extra particles are needed to support such symmetries.

In the minimal LRSM Dirac neutrino masses would now be subjected to obey similar patterns than those on the charge lepton sector, due to the common origin of mass terms, which usually arise from the same set of Yukawa couplings. This is of course a consequence of the fact that  in the LRSM, RHN's are paired with the right-handed leptons, which introduces by construction a meaningful lepton number for both the handed sectors. This makes the identification of $\mu\lra\tau$ symmetry neat.
The realization of flavor symmetries becomes more constrained and a bit more challenging from the theoretical point of view, though. In the particular case of $\mu\lra\tau$ symmetry, its realization requires the model to provide  diagonal charge lepton masses at tree level, from where different  $\mu$ and $\tau$ masses should arise naturally. This comes with the consequence of having a similar (diagonal) mass structure for the tree level Dirac neutrino masses, and thus, introducing an extra tree level source for the breaking of $\mu\lra\tau$ symmetry.  Yet, one would like to have small predictions for the value of $\theta_{13}$ and the deviation of maximality for $\theta_{23}$, at most closer to the current experimental limits. The smallness of  the induced breaking terms  would possibly be a matter of parameter tuning, though.
Another generic implication of these models would be that Dirac masses should not generate mixings. Lepton mixings would come solely from the Majorana mass terms, reducing in this way the number of free parameters that the theory would involve to reproduce the low energy data. To explore these aspects of such models is the main task of the present work.

$\mu\lra\tau$  flavor symmetry can be implemented in any model with the help of discrete groups like  $Z_{2}$, $S_{3}$, and $A_4$ among others \cite{numodels},
which still makes it possible to build different realizations of the  model with exact and broken $\mu\lra\tau$ symmetry . We will follow this route for our model with the goal of providing a framework where a small broken $\mu\lra\tau$ symmetry is realized and where charged lepton mass matrix
$m_{\ell}$ is given in a properly diagonal form at tree level.
Thus, we propose a LRSM with a $(Z_{2})^{3}$ set of discrete
flavor symmetries that provides the
realization of a slightly broken $\mu\lra\tau$ symmetry, broken only by the diagonal
Dirac mass terms at tree level,  with lepton mixings arising
from the (yet $\mu\lra\tau$ symmetric) Majorana mass matrix alone. ( For models that use similar discrete symmetries in the context of the
SM see, for instance Ref, \cite{grimus}.)
In this model, the breaking of the discrete symmetries
shall induce sizable non zero $\theta_{13}$, and the deviation of $\theta_{23}$ from $45^{\circ}$,
controlled by a single parameter associated with the Yukawa couplings on charged lepton and Dirac
neutrino masses.
The paper is organized as follows: first of all, we present our extended LRSM with discrete
symmetries and briefly argue on the scalar sector in Sec. II.
In Sec. III, we comment on the way we realize a small broken $\mu\lra\tau$ symmetry. We present, in Sec. IV, our predictions for the mixings with a minimal number of free
parameters, assuming no CP violation. Finally, we show the results obtained in the above section, in the presence of CP-violating phases, do not change drastically in general. We  close our discussion  with a summary of conclusions.

\section{The extended left-right model with $\mu\lra\tau$ symmetry}

The model we shall consider along our discussion is based on the usual
$SU(2)_L\times SU(2)_R\times U(1)_{B-L}$ gauge symmetry, but with an enlarged matter content
that includes two scalar
bidoublets $\phi_{i}(B-L = 0)$ ($i=1,2$),  and four $SU(2)_R$ triplets, $\Delta_{j}(B-L=-2)$
($j=1,...,4$), aside from the usual left $L_\alpha(B-L=-1)$ and right $R_\alpha(B-L=1)$ leptonic doublet representations, for $\alpha=e,\mu,$ and $\tau$. Here we assume parity conservation, where parity means the transformations: $L\leftrightarrow R$, $\phi_{i}\leftrightarrow\phi^{\dag}_{i}$. Thus, there would also be a similar set of
$SU(2)_L$ triplets. Nonetheless,  they would not be involved in the mass generation since they would
have zero vacuum expectation values (VEVs), and thus,  we will not write them down explicitly.

In order to get a neutrino sector that is compatible with
observations we will introduce three $Z_2$ discrete symmetries under which
the matter fields should transform as: \bea
Z_{2}&:&\phi_{2}\lra-\phi_{2};L_{\mu}\lra L_{\tau};R_{\mu}\lra
R_{\tau};\Delta_{2}\lra\Delta_{3}~;\nonumber\\
Z_{2}^{\prime}&:&\phi_{2}\lra-\phi_{2};L_{\mu}\lra-L_{\tau};R_{\mu}\lra-R_{\tau};\Delta_{2}\lra-\Delta_{3}~;\nonumber\\
Z_{2}^{\prime\prime}&:&L_{\mu}\lra-L_{\mu};R_{\mu}\lra-R_{\mu};\Delta_{3}\rightarrow -\Delta_{3};
\Delta_{4}\rightarrow-\Delta_{4}.\nonumber~\label{1.1}\eea
where other fields not indicated remain invariant.
Notice that $Z_2$ realizes  $\mu\lra\tau$ symmetry for the model in an effective way.
It mimics $\mu\lra\tau$ symmetry for the lepton sector, provided that $\phi_2$ is an odd field under such a symmetry.
In the broken phase, however, the model will provide a mass pattern
that will certainly correspond to a slightly broken $\mu\lra\tau$ symmetry, as we will
see. In addition, $Z_2^\prime$ and
$Z_2^{\prime\prime}$ are additional symmetries whose role is to cancel all off-diagonal couplings among charged leptons, and thus the appearance of a flavor-changing neutral current at tree level. The
above class of  symmetries
has been used before in the context of $SU(2)_{L}\times U(1)_{Y}$ models with three
$\nu_{R}$ right-handed neutrinos~\cite{grimus}, but in such a model $\mu\lra\tau$ symmetry remains exact, and
lepton mixings emerge from the Majorana mass matrix. In our case, we will rather obtain a framework where
this symmetry is broken at tree level in the effective neutrino mass
matrix, following the spontaneous breaking of symmetries. Some details regarding the analysis of the
scalar potential with this particle content and symmetries are given in the Appendix.

Next, according to the above symmetries, the most general Yukawa couplings of leptons to
Higgs bidoublets are given as \bea
\overline{L}_{e}\left[h\phi_{1}+\widetilde{h}\widetilde{\phi}_{1}
\right]R_{e}
+\overline{L}_{\mu}\left[h_{1}\phi_{1}+\widetilde{h}_{1}\widetilde{\phi}_{1}
- h_{2}\phi_{2}-\widetilde{h}_{2}\widetilde{\phi}_{2}
\right]R_{\mu}\nonumber\\
+\overline{L}_{\tau}\left[h_{1}\phi_{1}+\widetilde{h}_{1}\widetilde{\phi}_{1}
+ h_{2}\phi_{2}+\widetilde{h}_{2}\widetilde{\phi}_{2}
\right]R_{\tau}+ h.c.~~~\nonumber\eea
where $\widetilde{\phi}=\tau_{2}\phi^{\ast}\tau_{2}$.
Parity symmetry is explicitly shown in the above equation. This is a feature
of the LRSM before spontaneous symmetry breaking. On the other hand,
we can see that the combined discrete symmetries restrict enough above
couplings, in a way such that only diagonal mass terms  arise.
Then, we can identify our basis with the true flavor. As
is clear, after symmetry breaking one shall get the correct charged lepton masses
$m_{\mu}\neq m_{\tau}$, signaling the breaking of $\mu\lra\tau$
exchange invariance. By construction, such a breaking will also appear in the
Dirac neutrino mass terms, which shall follow a similar profile.
Moreover, this will be the only place where the
exchange symmetry is broken afterward.

The Lagrangian for the Yukawa couplings to triplets, allowed for the
symmetries, is written as
\bea
&\left[f_{1}\overline{(R_{e})^{C}}R_{e}
+f_{2}\left(\overline{(R_{\mu})^{C}}R_{\mu}+
\overline{(R_{\tau})^{C}}R_{\tau}\right)\right]\Delta_{1}\nonumber\\
&+f_{12}\bigg[\left(\overline{(R_{\mu})^{C}}\Delta_{3}+
\overline{(R_{\tau})^{C}}\Delta_{2}\right)R_{e}
+\overline{(R_{e})^{C}}\bigg(
\Delta_{3}R_{\mu}\nonumber\\&+\Delta_{2}R_{\tau}\bigg)\bigg]
+f_{23}\left[\overline{(R_{\mu})^{C}}\Delta_{4}R_{\tau}+
\overline{(R_{\tau})^{C}}\Delta_{4}R_{\mu}\right]\nonumber\\&+\left[R\rightarrow
L\right]+h.c.\label{1.3}\eea
Here, the four triplets help us build a
Majorana matrix which  generates lepton mixings.
As a matter of fact, since the Dirac mass matrix will be diagonal, the only source of mixings
should arise from Eq.(\ref{1.3}).
It is worth noticing the way the model separates the physical sources of mixings from which breaks $\mu\lra\tau$, although both components get finally interconnected by the see-saw at low energy.

To get masses for all fermions, we have to
break the gauge $SU(2)_{L}\times SU(2)_{R}\times U(1)_{B-L}$
symmetry in the usual way. So, the scalar field vacuum
expectation values are given as \be \langle\phi_{i}\rangle =
\left( \ba{cc}
k_{i}& 0\\
0& k^{'}_{i}\\
\ea \right)\quad\mbox{and}\quad
 \langle\Delta_{j} \rangle= \left( \ba{cc}
0 & 0\\
V_{j}& 0\\
\ea \right)~;\label{1.4}\ee
where, in order to have the see-saw mechanism~\cite{see-saw} at work,
we have to assume that right-handed triplet VEVs are larger than those of the bidoublets.
So far, for simplicity we are leaving out
CP-violating phases in our analysis, and so we neglected all  phases in the
vacuum expectation values,  as well as in Yukawa couplings. We do remark that
this assumption is motivated by the larger number of scalar fields in the model,
and it is taken in order to reduce the free parameters of the model to be
considered, in this case CP-violating phases. Nevertheless, in general, we do not
expect CP phases to drastically change the general predictions of the model.
Particularly, we see no indication that the order of magnitude of
$\sin\theta_{13}$ and of $1/2 - \sin^2\theta_{23}$ could be affected in a
significant manner, so we rather prefer to keep the analysis simple, since
estimating such numbers is the main goal for what follows. In addition, to give
enough support to our statement we will explore, later on, the model in the
presence of CP-violating phases in an effective way.

As we already mentioned, the
charged lepton mass matrix,  $m_{\ell}$, is diagonal
as well as  the Dirac
neutrino mass matrix, $m_{D}$. Their entries are
explicitly given as
\bea m_{e}&=&hk_{1}^{'}+\widetilde{h}k_{1},~
m_{\mu,\tau}=h_{1}k_{1}^{'}+\widetilde{h}_{1}k_{1}\mp(h_{2}
k_{2}^{'}+\widetilde{h}_{2}k_{2});\nonumber\\
\widetilde m_{e}&=&hk_{1}+\widetilde{h}k_{1}^{'},~
\widetilde m_{\mu,\tau}=h_{1}k_{1}+
 \widetilde{h}_{1}k_{1}^{'}\mp(h_{2}k_{2}+\widetilde{h}_{2}k_{2}^{'})\label{1.5}\nonumber\\
\eea
respectively,
where Dirac neutrino masses, $\widetilde m_{\ell}$, have been labeled according to the corresponding lepton number.
Notice that the model produces a $\mu$-$\tau$ mass difference in a natural way. Notice also the similar profile that arises in neutrino masses where
$\mu$-$\tau$ symmetry turns out to be broken by the essentially the same source. An interesting point that arises here is that the hierarchy between the last ($\widetilde{m}_{\mu}$ and $\widetilde{m}_{\tau}$) two Dirac masses is much softer than the ($m_{\mu}$ and $m_{\tau}$) charged lepton masses, as a result of Eq. (\ref{1.5}). Even if there were strong cancellations among the first and second terms in $m_\mu$, as required to get the right hierarchy of charged lepton masses, that would not be the same for
$\tilde m_\mu$, which allows for some level of degeneracy on Dirac neutrino masses, in accordance with $\mu\lra\tau$ symmetry. At this level
parity symmetry is broken since we have taken zero VEVs for the left-handed scalar triplets, and therefore, the type-I see-saw mechanism arises to explain  small neutrinos masses.

On the other hand, we do stress that the right-handed neutrino mass matrix, $M_R$, is symmetric under $\mu\lra\tau$.
This is due to the fact that the
vacuum expectation values  that
minimize the potential satisfy $\Delta_{2} =\Delta_{3}$ ($V_2 = V_3$).
This happens as a consequence of the discrete symmetries of the model, as we explicitly show in the Appendix. Thus, we get \be M_{R}=\left(\ba{ccc}
f_{1}V_{1} & f_{12}V_{2} &f_{12}V_{2}\\
f_{12}V_{2}& f_{2}V_{1}  &f_{23}V_{4}\\
f_{12}V_{2}& f_{23}V_{4} &f_{2}V_{1}
\ea\right)~;
\label{1.6} \ee
which clearly exhibits a $\mu$-$\tau$ symmetric structure.
It is worth mentioning that even though $M_R$ is the only source for neutrino mixings, by itself alone it does not fix $\theta_{23}$, nor $\theta_{13}$. It is the conspiracy with the Dirac mass terms that do the fixing.
For simplicity, we parametrize the inverse Majorana mass matrix as
\be M_{R}^{-1}=\left(\ba{ccc}
x & y & y\\
y & w & z\\
y & z & w
\ea\right)~,
\label{1.6.1}\ee
which of course maintains a symmetric profile. These new  parameters  are easily rewritten in terms of the $M_R$ elements, but for the low energy analysis it is enough to use them instead, without obscuring the physical meaning.

We ought to comment on two important things about the model: first of all, we should emphasize that, after  spontaneous symmetry  breaking, the  Dirac mass terms,
$\overline{\ell}_{L}m_{\ell}\ell_{R}+ \overline{\nu}_{\ell L} m_{D}\nu_{\ell R}$, are not any more invariant under $\mu\lra\tau$ because of the non zero mass differences $m_{\tau}- m_{\mu}$ and
$\widetilde m_{\tau}- \widetilde m_{\mu}$,  as we can see from Eq. (\ref{1.5}), whereas the Majorana mass
terms do possess exact $\mu\lra\tau$ symmetry. Second, the  symmetry breaking in the diagonal Dirac mass matrix
will generate the same effect in the effective neutrino mass matrix, $M_\nu$, when the see-saw mechanic is implemented. Indeed, one gets the mass matrix
\be
 M_{\nu}=-\left(\ba{ccc}
 m_{ee} & m_{e\mu}&m_{e\tau}   \\[.5ex]
 m_{e\mu}& m_{\mu\mu}&m_{\mu\tau}\\[.5ex]
 m_{e\tau}& m_{\mu\tau}&m_{\tau\tau}
 \ea\right),
\label{1.7} \ee
that exhibits the breaking of the   $\mu \lra\tau$ symmetry at tree level through the mass differences
$m_{e\tau}-m_{e\mu}=y\widetilde m_{e}(\widetilde m_{\tau}-\widetilde m_{\mu})\neq0$ and $m_{\tau\tau}-m_{\mu\mu}=w(\widetilde m_{\tau}^{2}-\widetilde m_{\mu}^{2})\neq0$.

Although, this symmetry is not exact in the neutrinos sector, it can be considered as an approximate symmetry as a result of a small breaking in the mass difference $\widetilde{m}_{\tau}-\widetilde{m}_{\mu}$. It is allowed, as we show later, because we have the freedom to manipulate the free parameters of the LRSM, specifically in Eq. (\ref{1.5}). The effective mass terms can be separated as $\overline{\nu}_{L}(
M_{\nu})_{\mu\lra\tau}\nu_{L}^{C}+\overline{\nu}_{L}\delta M_{\nu}\nu_{L}^{C}$,
where the last term contains explicitly the breaking. Such a
term  affects lepton mixings, inducing  calculable deviations
in the mixing angles from  the bimaximal mixing scenario.

\section{Masses and mixings}
The matrix $M_{\nu}$ is diagonalized by
$U_{PMNS}$ (in the standard parametrization), with
$\theta_{23}\approx \alpha-\frac{\pi}{4}$ where $|\alpha| \ll 1$. Notice that the $\alpha$ parameter can lay in the first or fourth quadrant of parameter space.
$\theta_{12}$ is then identified as $\theta_{\odot}$ and now $\theta_{13}$ should be non zero. When
$\theta_{13}$ and $\alpha$ go to zero, the mixing matrix becomes that of the
bimaximal framework. Under this approximation, the following  three conditions should be fulfilled in order to get
a diagonal mass matrix:
\bea \tan2\theta_{\odot}&\approx&
\sqrt{8}\left[\frac{\overline{m}_{e\mu}}{\overline{m}_{\mu\mu}+\left(m_{\mu\tau}-m_{ee}\right)}\right];~~~\label{1.8}\\
\sin\theta_{13}
&\approx&\frac{1}{\sqrt{8}}\left[\frac{2m_{\mu\tau}\delta-\epsilon\overline{m}_{e\mu}}
{\overline{m}_{e\mu}^{2}+m_{\mu\tau}\left(\overline{m}_{\mu\mu}-m_{\mu\tau}-m_{ee}\right)}
\right];~~~\label{1.9}\\
\sin\alpha
&\approx&-\frac{1}{4}\left[\frac{\epsilon\left(\overline{m}_{\mu\mu}-m_{\mu\tau}-m_{ee}\right)+2\delta\overline{m}_{e\mu}}
{\overline{m}_{e\mu}^{2}+m_{\mu\tau}\left(\overline{m}_{\mu\mu}-m_{\mu\tau}-m_{ee}\right)}
\right],~~~\label{1.10}\eea
where we have defined
$\overline{m}_{e\mu}=\frac{1}{2}(m_{e\mu}+m_{e\tau})$, $\overline{m}_{\mu\mu}=\frac{1}{2}(m_{\tau\tau}+m_{\mu\mu})$, $\delta=m_{e\tau}-m_{e\mu}$ and $\epsilon=m_{\tau\tau}-m_{\mu\mu}$.
The last two parameters give us an account of how strong the breaking of the $\mu\lra\tau$
symmetry is. We expect them to be small, compared to $m_{e\mu}$ and
$m_{\mu\mu}$, respectively, since they
are related  to $\alpha$ and $\theta_{13}$ as we can see from Eqs. (\ref{1.9}) and (\ref{1.10}).
On the other hand,
the eigenvalues of $M_{\nu}$ are given as
\bea m_{1}&\approx&m_{ee}-\sqrt{2}\overline{m}_{e\mu}\tan\theta_{\odot};\label{1.11}\\
m_{2}&\approx&m_{ee}+\sqrt{2}\overline{m}_{e\mu}\cot\theta_{\odot};\label{1.12}\\
m_{3}&\approx&\overline{m}_{\mu\mu}-m_{\mu\tau}.\label{1.13}
\eea

Next, we would like to get some predictions out of  this model,
particularly for Eqs. ($\ref{1.9}$) and ($\ref{1.10}$) that depend on the
elements of $M_{\nu}$.
In order to do this, we have to identify and properly choose the free parameters of our model. As we can see in $M_{\nu}$, if one writes this matrix and Eqs. (\ref{1.8})-(\ref{1.10}), in terms of Dirac and Majorana masses, one realizes that there are seven parameters: $x$, $y$, $w$, and $z$, which  stand for the elements of the inverse matrix $M_{R}^{-1}$, and $\widetilde{m}_{e, \mu, \tau}$ corresponding to Dirac masses. We would like now to fix some of those parameters using neutrino observables, others than those we are intending to predict at some level. Thus we take $\Delta m_{\odot}^{2}=m_{2}^{2}-m_{1}^{2}$,
$\Delta m_{ATM}^{2}=\frac{1}{2}|m_{1}^{2}+m_{2}^{2}-2m_{3}^{2}|$, the scale of neutrino mass hierarchy $m_{3}$, and the solar
mixing angle $\theta_{\odot}$ as our inputs to fix the free parameters as much as possible.
In spite of the lack of observables, we can only  determine four of the model free parameters in terms of these known ones. So, we decided to fix $x$, $y$, $w$, and $z$, keeping in mind that the Dirac neutrino mass matrix is not invariant under $\mu\lra\tau$ symmetry, and, as we have commented, that this is the only source of symmetry breaking. Thus, from the above  given definitions of $\Delta m_{\odot}^{2}$, $\Delta m_{ATM}^{2}$, and using Eqs. (\ref{1.8}) and (\ref{1.13}), we can solve the system of variables for the four inverse  Majorana masses,  which will be now given as
\bea
x&=&\frac{1}{\widetilde{m}^{2}_{e}}\left[\sin^{2}\theta_{\odot}A+\cos^{2}\theta_{\odot}B\right];\nonumber\\
y&=&\frac{\sin2\theta_{\odot}}{\sqrt{2}\widetilde{m}_{e}\left(\widetilde{m}_{\tau}+\widetilde{m}_{\mu}\right)}\left[A-B\right];\nonumber\\
w&=&\frac{1}{\left(\widetilde{m}^{2}_{\tau}+\widetilde{m}^{2}_{\mu}\right)}\left[\sin^{2}\theta_{\odot}B+\cos^{2}\theta_{\odot}A+m_{3}\right];\nonumber\\
z&=&\frac{1}{2\widetilde{m}_{\tau}\widetilde{m}_{\mu}}\left[\sin^{2}\theta_{\odot}B+\cos^{2}\theta_{\odot}A-m_{3}\right],
\label{1.17}
\eea
where
\bea
A&=&\sqrt{m_{3}^{2}\pm \Delta m_{ATM}^{2}+\frac{1}{2}\Delta m_{\odot}^{2}}~; \nonumber \\
B&=&\sqrt{m_{3}^{2}\pm \Delta
m_{ATM}^{2}-\frac{1}{2}\Delta m_{\odot}^{2}}~;
\eea
and  $+$ ($-$) stands for the inverted (normal) hierarchy case. These expressions would allow us to simplify our predictions for the mixings to write them in terms of neutrino observables, as we shall see next.
So far, we still have three (Dirac mass) free parameters but, as it is easy to realize,
the first Dirac mass, $\widetilde m_e$, shall not be involved in the  determination of  $\sin\theta_{13}$ and $\sin\alpha$, since it has a null contribution to  $M_{\nu}$ at the lower seesaw order. Therefore, as far as it matters for the mixings, we end up with two unknown parameters,  which can be reduced to one by further considerations, as we will comment on next. Hence, by introducing the above expressions for the four Majorana masses  into Eqs. (\ref{1.9}) and (\ref{1.10}), one gets the  analytical forms
\scriptsize
\bea
\sin\theta_{13}\approx&-\frac{s2\theta(A-B)}{4}\left[\frac{(2r-R)(Ac^{2}\theta+Bs^{2}\theta)-m_{3}(2r+R)}{m_{3}^{2}-m_{3}(A+B)+AB}\right];\label{1.18}\\
\sin\alpha\approx&\left[\frac{rs^{2}2\theta(A-B)^{2}-2R(Ac^{2}\theta+Bs^{2}\theta+m_{3})(As^{2}\theta+Bc^{2}\theta-m_{3})}{4\left(m_{3}^{2}-m_{3}(A+B)+AB\right)}\right],\label{1.19}\eea
\normalsize
where we have defined $r=(\widetilde{m}_{\tau}-\widetilde{m}_{\mu})/(\widetilde{m}_{\tau}+\widetilde{m}_{\mu})$ and $R=(\widetilde{m}^{2}_{\tau}-\widetilde{m}^{2}_{\mu})/(\widetilde{m}^{2}_{\tau}+\widetilde{m}^{2}_{\mu})$. Also, in these formulas, we have used a simplifying notation where $s^2\theta$, $c^2\theta$, and $s2\theta$ ($s^22\theta$) stand for $\sin^2\theta_{\odot}$, $\cos^2\theta_{\odot}$ and $\sin2\theta_{\odot}$ ($\sin^22\theta_{\odot}$), respectively.
It is worth remarking how interesting a feature the parametric $\widetilde{m}_{e}$ Dirac mass independence of $M_\nu$ becomes. Because of this fact, $m_{ee}$, which fixes the strength of double beta decay, is very well determined by the observables, as one can notice from the first expression in Eq. (\ref{1.17}), which gives
\be
m_{ee}=  \sin^{2}\theta_{\odot}A+\cos^{2}\theta_{\odot}B~,
\ee
that only depends on the unknown mass scale $m_3$ and the hierarchy.

In addition, we have two ($\widetilde{m}_{\tau}$ and $\widetilde{m}_{\mu}$) free parameters whose difference is the only source of $\mu\lra\tau$ symmetry breaking. Given the  ratio forms  $r$ and $R$, we can introduce a further phenomenologically motivated reduction of the parameters, by observing that $\widetilde{m}_{\tau}-\widetilde{m}_{\mu}$ must be small to truly generate a tiny breaking of the symmetry. In order to do that, let us take
the relation between charged lepton and Dirac masses from Eq.(\ref{1.5}).
Thus, without loss of generality, we can take
$k_{i}^{'}\ll k_{i}$ and assume that Yukawa couplings $h_{1}$ and $h_{2}$ are smaller than $\widetilde{h}_{1}$ and $\widetilde{h}_{2}$, respectively. Then, we easily imply that the
parameter $h_{2}/\widetilde{h}_{2}$ should be small to generate a tiny  symmetry breaking. However, knowing that $h_{1}<\widetilde{h}_{1}$, its ratio will be further constrained for the condition $h_{s}=\frac{h_{2}}{\widetilde{h}_{2}}/\frac{h_{1}}{\widetilde{h}_{1}}\ll1$, and  as a result of the definition of $r$ and $R$ they will appear to be only dependant on the new parameter $h_s$. Therefore, with this simple approximations we can reduce the number of effective free parameters from seven to five.  It is worth stressing that this is possible since the final expressions for $\sin\theta_{13}$ and $\sin\alpha$ do not  depend on Dirac neutrino mass but throughout the mass ratios $r$ and $R$, which can finally be simplified by  hiding one of the parameters in an effective ratio, as we have seen.

\section{Model predictions}
As a result of the approximations introduced above, the expressions for $\sin\theta_{13}$ and $\sin\alpha$
given in Eqs.(\ref{1.18}-\ref{1.19}) are  simplified into
\bea
\sin\theta_{13}&\approx&m_{3}h_{s}\overline{m}\left[\frac{\sin2\theta_{\odot}
\left(A-B\right)}{m_{3}^{2}-m_{3}\left(A+B\right)+AB}\right];\label{5.14}\\
\sin\alpha&\approx&h_{s}\overline{m}
\left[\frac{m_{3}^{2}+m_{3}\cos2\theta_{\odot}
\left(A-B\right)-AB}{m_{3}^{2}-m_{3}\left(A+B\right)+AB}\right]\label{1.20},
\eea
where $\overline{m}=(m_{\tau}-m_{\mu})/(m_{\tau}+m_{\mu})$.
We note that these expressions have a linear dependence on $h_{s}$, and this will allow us to get a direct correlation between the angles that do not depend on $h_{s}$,  as we will show later on. Next, although we could obtain explicit expressions for each hierarchical case,
we rather prefer to explore a more pictorial representation of the  parameter space, by showing the experimentally allowed regions for $h_{s}$ and the hierarchy mass scale $m_{3}$, such that $\sin\theta_{13}$ and $\sin\alpha$ are right below the current experimental limits. To be more explicit, we consider the upper limit value for $\sin\theta_{13}$ to be of the level of $1\sigma$ deviation, as given by the fits of current neutrino experiments data~\cite{valle1,theta13}. We take the upper limit to be $\sin\theta_{13}<0.16$ for the sake of demonstration. Additionally, we observe that  $|\sin\alpha|\approx|\frac{1}{2}-\sin^{2}\theta_{23}|$,
and thus $\sin\alpha$ can be considered to be at most of the level of $1\sigma$ deviation from  the central value in $\sin^2\theta_{23}$, such that, at the end of the day we are  indirectly measuring $\sin\theta_{23}$. Here we assume  the upper value $|\sin\alpha|<0.07$. Under these two premises, we can plot the corresponding $\sin\theta_{13}$ and $\sin\alpha$ contour values in the $m_3-h_s$ parameter space, respectively, that should bound the allowed parameter space.
Our results are depicted in figures 1 and 2.

\begin{figure}[ht]
\begin{center}
\includegraphics[width=180pt]{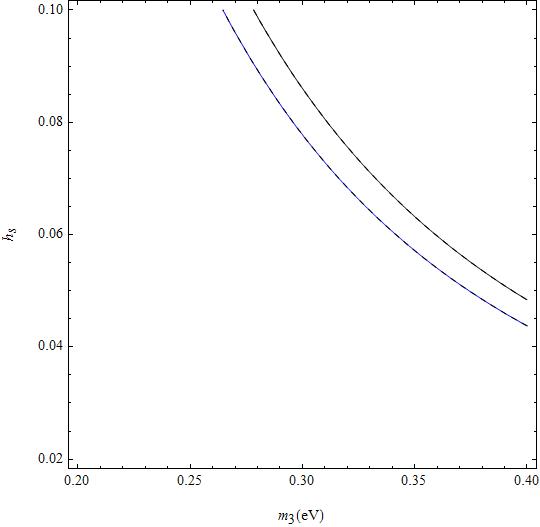}\\
\caption{The $1\sigma$ region for
$\sin\theta_{13}$ in the parameter space of $h_s$ and $m_{3}$, for both  normal (solid lines)
and inverted (dashed lines) hierarchies. The upper and lower contours have a value of $\sin\theta_{13}=0.16$, within a $\sigma$ deviation. Notice that solid and dashed lines completely overlap}\label{sinteta}
\end{center}
\end{figure}

\begin{figure}[ht]
\begin{center}
\includegraphics[width=180pt]{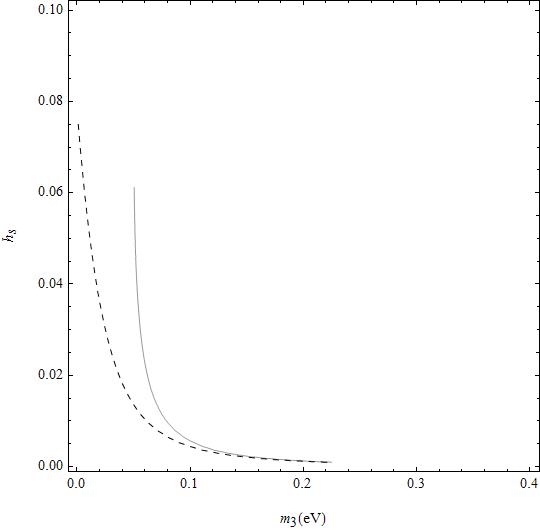}\\
\caption{The $1\sigma$ region for
$\sin\alpha$ as a function of $h_{s}$ and $m_{3}$, for both  normal (solid line)
and inverted (dashed line) hierarchies. The upper and lower contours correspond to the limit value of $\sin\alpha=0.07$.}\label{sinalpha}
\end{center}
\end{figure}

{}From Fig. \ref{sinteta}, for $\sin\theta_{13}$ all the region below the curves is physically
acceptable, within $1\sigma$ deviation. At the same time, one observes that the $\sin\theta_{13}$ allowed region is mostly insensitive to hierarchy, meaning,  we can not differentiate between normal and inverted hierarchy. Indeed, the regions basically overlap, so that, both hierarchies share most of the same parameter region. In particular, two extreme points we have to comment on: for small values of $h_{s}$, $m_{3}$ is close to its maximal value. On the other hand, for large values of $h_{s}$, $m_{3}$ is larger than its middle maximal value. In addition, from the figure one may identify a window of acceptable values, which lies within  $h_{s}\approx\left(0.04-0.1\right)$ and $m_{3}\approx\left(0.26-0.4\right)$. The important result here is that $h_{s}$ must be tiny in order to have a consistent value for $\sin\theta_{13}$, according to the experimental limits. The same conclusion will arise for $\sin\alpha$, as we will see from Fig.~\ref{sinalpha}.

Before starting the discussion about Fig.~\ref{sinalpha}, we would like to comment that we have taken the absolute value of $\sin\alpha$ just for the sake of depicting together all cases in the same plot. However, the case of inverted hierarchy  produces negative values for $\sin\alpha$, which already introduces a way to distinguish the hierarchy if one could identify in what direction $\theta_{23}$ deviates from the maximal value. We also observe that normal and inverted hierarchy can easily differentiate at small values of $m_{3}$. However, for tiny values of $h_{s}$ (and large values for $m_{3}$) both of them are  difficult to point out from each other (apart from the overall sign). This plot is more constrained since the allowed parameter region for $\sin\alpha$ remains below  the contours for each hierarchy. Again, from this figure we infer the phenomenologically valid interval of values that gets within $h_{s}\approx\left(0.001-0.07\right)$ and $m_{3}\approx\left(0.001-0.22\right)$ for inverted hierarchy, and $h_{s}\approx\left(0.001-0.06\right)$ and $m_{3}\approx\left(0.06-0.22\right)$ for normal hierarchy. Comparing the parameter regions of $m_{3}$ for $\sin\theta_{13}$ and $\sin\alpha$ graphics, this seems to be in disagreement since both sets of values should be common in order to satisfy simultaneously such expressions. Nonetheless, in Fig. 3 we show the valid region of values for $m_{3}$ to both hierarchy cases.

\begin{figure}[ht]
\begin{center}
\includegraphics[width=180pt]{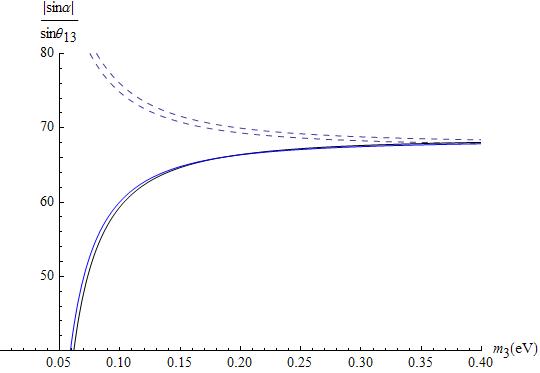}\\
\caption{The $1\sigma$ region for
$|\sin\alpha|/\sin\theta_{13}$ as function of $m_{3}$, for both  normal ( solid lines)
and inverted ( dashed lines ) hierarchies.}\label{ratio}
\end{center}
\end{figure}

One interesting prediction of the model is that the ratio $\sin\alpha/\sin\theta_{13}$ does not depend on the  parameter $h_{s}$, as we already pointed out, and thus, an appealing formula arises which represents the key expression that allows one to falsify this model
\be
\frac{\sin\alpha}{\sin\theta_{13}}\approx\left[\cot2\theta_{\odot}+\frac{m^{2}_{3}-AB}
{m_{3}\sin2\theta_{\odot}(A-B)}\right],\label{1.21}
\ee
This ratio is principally driven by the second term which has a clear dependence on $m_{3}$ and the rest of the observables, and
also an extra contribution comes from the solar angle.
This can be enhanced or decreased depending on the hierarchy scheme, in particular, for strict normal hierarchy one obtains a fit value of about $68.45$ when  central values are taken for the observables. On the other hand, inverted hierarchy has a notable dependence on $m_{3}$ whose value is not well determined. One can in general depict the expected  ratio in terms of $m_3$, as we show explicitly in Fig.~\ref{ratio}.

As we can see from Fig.~\ref{ratio}, there is a clear difference between both hierarchies for small values of the  $m_{3}$ mass scale.
For normal hierarchy the ratio decreases with the hierarchy scale $m_{3}$, whereas for inverted hierarchy the ratio rather increases (negatively) with $m_3$. However, it is hard to differentiate between the two hierarchies for large $m_3$. The most important point we do want to make about this plot is that our model, under the previous approximations, predicts that due to the  correlation between $|\sin\alpha|$ and $\sin\theta_{13}$,  the latter parameter will very likely lie in below a range of $10^{-3}$, which can be difficult to reach in  future experiments, given that $|\sin\alpha|\approx|\frac{1}{2}-\sin^{2}\theta_{23}|$ is already bounded below $10^{-2}$ by current  experimental data. Nevertheless, this correlation will make it possible to falsify our model in the following way. If future experiments achieve to measure $|\sin\alpha|$ and/or $\sin\theta_{13}$,  then, they should lie in the allowed region according to Fig.~\ref{ratio}. On the contrary, if such were not the case, the model would be automatically discarded.

\section{CP phase contribution}

Next, let us add some comments on CP phase contributions. As we already stated,
and as we will discuss below, it seems unlikely that such contributions may
drastically change our results. In particular, aside from some small numerical
changes, it does appear yet difficult that CP phases may  increase the tight
upper bound the model indicates for $\sin\theta_{13}$, which has been put on at
about $10^{-3}$, and which is already far away
from the current experimental bound ($\sim 10^{-1}$).

In order to show our present point in a simple way,
let us briefly reconsider our analysis of Sec. III.
As usual, the mixing matrix is in general given by $U=U_{PMNS}K$,
where $K= diag(1,e^{i\phi_{1}},e^{i\phi_{2}})$ stand for the Majorana CP phases.
Also, $U_{PMNS}$ contains (in the standard parametrization) the $\varphi$ Dirac
CP phase. So that, under the same considerations as for Sec. III, $M_{\nu}$
is diagonalized by $U$, and Eqs.(\ref{1.9}-\ref{1.10}) get replaced by
\bea
z&=&\frac{1}{\sqrt{2}}\left[\frac{P^{\ast}\left(\delta-\epsilon
C\right)-K\left(\delta-\epsilon C\right)^{\ast}}{\mid P\mid^{2}-\mid
K\mid^{2}}\right];\label{newsintheta}\\
\sin\alpha&=&-\frac{1}{4m_{\mu\tau}}\left[\epsilon+\sqrt{8}z\overline{m}_{e\mu}
\right]\label{newsinalpa}
\eea
where, as before, $z=\sin\theta_{13}e^{i\varphi}$, and we have introduced the
short-hand notation where
$P=\overline{m}^{2}_{e\mu}/m_{\mu\tau}-m_{ee}$,
$K=\overline{m}_{\mu\mu}-m_{\mu\tau}$ and
$C=\overline{m}_{e\mu}/2m_{\mu\tau}$. Here, $\delta$ and $\epsilon$
parameters have the same form as in previous sections. We must keep in mind the
$M_{\nu}$ matrix elements are now complex ($\mid m_{\alpha\beta}\mid
e^{i\rho_{j}}$, where $\alpha,\beta=e,\mu, \tau$, and $j=1,..,6$), and the
Majorana CP phases are implicitly included in Eqs.
(\ref{newsintheta}-\ref{newsinalpa}). Of course, when
CP phases are canceled (taking all masses to be real), the above expressions
reduce to the previous results as given in Eqs. (\ref{1.9}-\ref{1.10}).
Interesting enough, Eq. (\ref{1.8}) maintains its form even in the presence of
CP phases.

Having written the modified formulas for $\sin\theta_{13}$ and
$\sin\alpha$, we are now interested in  figuring out how  CP-violating
phases affect our previous considerations and conclusions. Performing a
complete and general analysis of the CP-violating case is challenging and
complicated. Nevertheless, there is a simpler
way to proceed for our purpose, and that is to use the numerical results that
came from our previous analysis as a grounding point, to estimate, through
a general example, the effect of  reintroducing non-trivial CP phases.
Hence, we will proceed as follows:
first, we use our CP-conserving results
to rebuild the effective neutrino mass matrix elements, thus identified as
$|m_{\alpha\beta}|$, for a given set of initial inputs,  and then
reintroducing CP phases on $M_\nu$, and  recalculating our predictions to
determine how far the new results are from those expected from the CP-conserving
case. We will particularly focus on the correlation given in
Eq. (\ref{1.21}), which is our main result here.
Moreover, since the absolute values of $\delta/m_{e\mu}$ and
$\epsilon/m_{\mu\mu}$ are supposed to be small,  we may conclude that
$m_{e\mu}$ and  $ m_{e\tau}$ will have about the same phase, as so will
$m_{\mu\mu}$ and  $ m_{\tau\tau}$, irrespective of $\delta$ and $\epsilon$
own arbitrary phases.
For simplicity, we will  suppose that $\overline{m}_{e\mu}$ and
$\delta$  have the same phase, and analogously for $\overline{m}_{\mu\mu}$ and
$\epsilon$; this choice shall
reduce the number of CP phases on $M_{\nu}$ to  four, although, it is clear
not all will be truly physical.
Under these considerations, the correlation between $\sin\alpha$
and $\sin\theta_{13}$ is now expressed as
\be
\frac{|\sin\alpha|}{|z|}\approx\frac{1}{\sqrt{2}|m_{\mu\tau}m_{ee}|}
\frac{\left| |\overline{m}_{\mu\mu}| f_ {1}
e^{i\gamma_{1}}+|\overline{m}^{2}_{e\mu}|f_{2}\right|}{\mid f_{2}\mid}
\label{newration}
\ee
where we have defined
\bea
f_{1}&=&
\left| |\overline{m}^{2}_{e\mu}|-|m_{\mu\tau}m_{ee}|e^{i\gamma}\right|^{2
}
\nonumber \\ & &
- \mid m_{\mu\tau} ^{2}\mid
\left| |\overline{m}_{\mu\mu}|e^{i\gamma_{1}}-|m_{\mu\tau}|\right|^{2};
\eea
and
\bea
f_{2}&=&\left[|\overline{m}_{e\mu}^{2}|-|m_{\mu\tau}m_{ee}|e^{-i\gamma}+
|m_{\mu\tau}^2|-|m_{\mu\tau}\overline{m}_{\mu\mu}|e^{-i\gamma_{1}}\right]
\nonumber \\  & &
\times \left(|m_{\mu\tau}|-|\overline{m}_{\mu\mu}|e^{i\gamma_{1}}\right).
\eea
Here, $\gamma= \rho_{1}+\rho_{3}-2\rho_{2}$ and $\gamma_{1}=\rho_{4}-\rho_{3}$
stand for the two effective CP-violating phases entering in our results that
came out of  combinations of the arbitrary phases we first introduced in the
neutrino mass matrix.
Therefore, we  realize the correlation among mixings depends only on  these two
CP phases, apart from the $m_{3}$ hierarchy mass scale.

\begin{figure}[ht]
\begin{center}
\includegraphics[width=180pt]{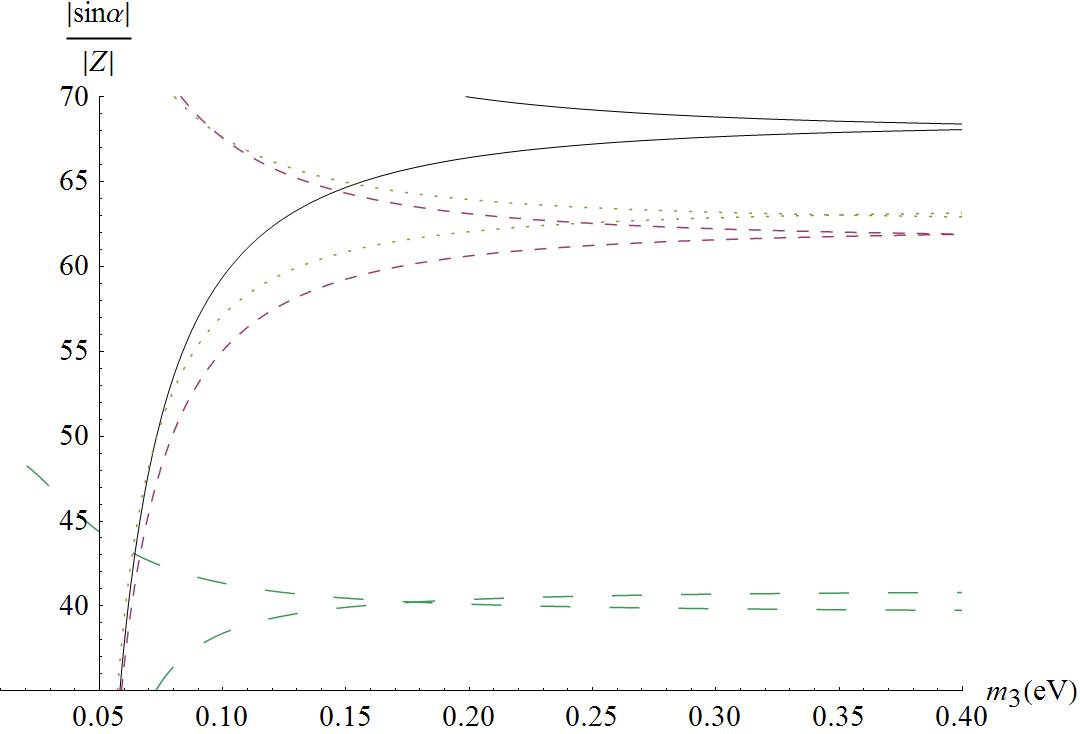}\\
\caption{Upper bound for
$|\sin\alpha|/|z|$ as a function of $m_{3}$, for different values of $\gamma$ and $\gamma_{1}$ as discussed in the text. As before, for normal
hierarchy, the contours (the correlation) decrease for small values of $m_{3}$
whereas for inverted hierarchy the contours increase
(negatively).}\label{ratio1}
\end{center}
\end{figure}

Next, we can use Eq.~(\ref{newration}) to quantify the phase dependence of
our previous CP invariant results. This we  do by exploring some general
examples, by taking specific values for $\gamma$ and $\gamma_{1}$, and then
replotting the ratio
$|\sin\alpha|/|z|$ as functions of $m_3$ for the central values of the observables. We show in
Fig~\ref{ratio1} (solid line) the results we obtained for the CP
invariant case, that is for $\gamma=0$ and $\gamma_{1}=0$, for both neutrino
mass hierarchies, as before
(see figure~\ref{ratio} for a comparison). Next, in same figure, short-dashed
lines have been obtained when the arbitrary values $\gamma=\pi/4$ and
$\gamma_{1}=\pi/5$ are taken. Similarly, for
$\gamma=\pi/2$ and $\gamma_{1}=\pi/4$ the associated plot is represented by the
dotted line. Finally, the long-dashed line stands for the arbitrary
set of values $\gamma=\pi/7$ and $\gamma_{1}=\pi/2$.

{}From Fig. (\ref{ratio1}), one may deduce important facts about the CP-
violating phases contributions to the correlation between $\sin\alpha$ and
$\sin\theta_{13}$: In general, for both hierarchies, CP phases do not change
the general profile of the $1\sigma$ regions, and neither seem to
introduce drastic changes on the $\sin\alpha/\sin\theta_{13}$ ratio, which
at least for the arbitrary values we have explored, keeps  $\sin\theta_{13}$
upper bound on  about $10^{-3}$, as we have pointed out before.
Indeed, in spite of
considering the largest changes we found on the  mixing
correlation (for $\gamma=\pi/7$ and $\gamma_{1}=\pi/2$),  the result
is not yet  enough to significantly reduce the bound.

Nonetheless, as a cautionary word, as we have not presented here a truly general
analysis for the CP-violating case, we can not discard the possibility that CP
phases may conspire among themselves to produce an accidental enhancement of the
$\sin\alpha/\sin\theta_{13}$ ratio.  A naive analysis shows that this
seems to happen for instance for the precise values of $\gamma=\pi$ and
$\gamma_{1}=\pi$, which set the mentioned ratio to the order of $10^{-1}$,
which in turn would locate $\sin\theta_{13}$ close to the experimental
sensitivity level. However, such a case has almost no parameter space on the CP
phase sector, and it seems very unlikely that such extreme values could be
potentially physical, since even phases are  subject to higher order
corrections, and no symmetry exists that may protect phases as to
keep them at those precise values.
Thus, a conservative point of view would point towards the strength of our
previous CP invariant results. A more careful and general study of  CP violation
in $\mu\lra\tau$ symmetric models is in progress,  and a detailed analysis of
this scenario will be presented soon elsewhere.

\section{Conclusions}
We studied the lepton sector of the LRSM dressed with a $(Z_{2})^{3}$ discrete symmetry, as was done in two frameworks: (1) neglecting the CP- violating phases and (2) involving them. Using discrete symmetries we build a framework where the charged lepton mass matrix is diagonal at tree level, so that we start since the beginning in the true flavor basis, and get a model that naturally realizes a slightly broken $\mu\leftrightarrow\tau$ symmetry. In addition, we have determined
$\theta_{13}$ and the deviation from the maximal value of $\theta_{23}$
in the limit of  a small $\mu\lra\tau$ symmetry breaking that appears at
tree level in the effective neutrino mass matrix. These two expressions
are fixed in terms of one free parameter of the model, $h_{s}$, the mass scale hierarchy $m_{3}$ (for the first scenario), and two CP-violating phases (for the second one); within the first scenario, a set of values for $h_{s}$ and $m_{3}$ were found which are consistent with current limits imposed by the experimental data for $\sin\theta_{23}$ and $\sin\theta_{13}$.

The present model is quite elaborate, but predictive. In particular, a
correlation among $\sin\alpha$ that measures the deviation of
$\sin^2\theta_{23}$ from $1/2$,  and $\sin\theta_{13}$ arises which only
depends on one neutrino mass scale, $m_3$ (and the CP phases). This is a
remarkable result and  provides a way to test the model, since
the observed value of $|\sin\alpha|/\sin\theta_{13}$ would also provide a direct
determination of the scale of $m_{3}$ for a given hierarchy.
We should point out this ratio does predict an allowed region for
$\sin\theta_{13}$ which must be below $10^{-3}$, in order to
satisfy the current experimental data, that upper bounds  the expression $1/2 -
\sin^2\theta_{23}$. Such a sensitivity could be hard to reach in future
experiments but it seems  possible. On the other hand, as an extra bonus of this
model, a clear determination of the $m_{ee}$ strength of double beta decay was
found.


\acknowledgments This work was supported in part by CONACyT,
M\'exico, grant No. 54576.

\section{appendix}
The most general form of the scalar potential of the LRSM was
analyzed in \cite{boris}, and it has 18 terms which are invariant under the gauge group $SU(2)_{L}\times SU(2)_{R}\times
U(1)_{B-L}$ and parity. Now, from our scalar particle content: two bidoublets $\phi_{i}$ ($i=1,2$), and four triplets,
$\Delta_{j}$ ($j=1,...,4$), we should write the potential. Even though,
it is larger than the original potential, it is possible to show by a lengthy calculation that it has a
least a minimum. Here we are interested in showing that the triplets
$\Delta_{2R}$ and $\Delta_{3R}$ have the same VEV when we minimize
the potential. So, without loss of generality, we just write the
terms in which their VEV'-s are involved after spontaneous symmetry breaking and they are given as:
\bea V&=&
\left(v_{2R}^{2}+v_{3R}^{2}\right)\bigg[-\mu_{1}^{2}+2bv_{1R}^{2}+2b_{1}v_{4R}^{2}+4b_{3}k_{1}k_{1}^{'}
\nonumber\\
&&+b_{2}\left(k_{1}^{2}+k_{1}^{'2}\right)+b_{4}k_{1}^{'2}+b_{5}\left(k_{2}^{2}+k_{2}^{'2}\right)
+4b_{6}k_{2}k_{2}^{'}\nonumber\\
&&+b_{7}k_{2}^{'2}\bigg]
+b_{8}\left(v_{2R}^{4}+v_{3R}^{4}\right)+2b_{9}v_{2R}^{2}v_{3R}^{2}
+...\eea
We can easily conclude from the above expression that $\Delta_{2R}$
and $\Delta_{3R}$ have the same VEV, by directly calculating the two
minimization conditions,
$\frac{\partial V}{\partial v_{2R}}=\frac{\partial V}{\partial v_{2R}}=0$.
It is not difficult to realize that this is a consequence of the fact that $V$ remains symmetric under the $v_{2R}\lra v_{3R}$ exchange, derived from the original $Z_2$ symmetry.
This ensures that the Majorana mass matrix,
$M_{R}$, will remain invariant under $\mu\lra\tau$ symmetry after spontaneous symmetry breaking. On the other hand, we have to keep in mind that
all VEV'-s will only depend on potential parameters  after solving
the simultaneous conditions for each VEV.

\end{document}